\documentclass[doublecol]{epl2} 
\title{Record-breaking temperatures reveal a warming climate}
\shorttitle{Record-breaking temperatures reveal a warming climate} %Insert here a short version of the title if it exceeds 70 characters

\author{G. Wergen\inst{1} \and J. Krug\inst{1}}
\shortauthor{G. Wergen \etal}

\institute{\inst{1}Institute for Theoretical Physics, University of Cologne, 50937 K\"oln, Germany}

\abstract{We present a mathematical analysis of records drawn from independent random variables with a drifting mean. To leading order the change in the record rate is proportional to the ratio of the drift velocity to the standard deviation of the underlying distribution. We apply the theory to time series of daily temperatures for given calendar days, obtained from historical climate recordings of European and American weather stations as well as re-analysis data. We conclude that the change in the mean temperature has increased the rate of record breaking events in a moderate but significant way: For the European station data covering the time period 1976-2005, we find that about 5 of the 17 high temperature records observed on average in 2005 can be attributed to the warming climate.}

\begin{document}

\maketitle

\section{Introduction}
In current media coverage the occurrence of record-breaking temperatures and other extreme weather conditions is often associated with global climate change. However, record breaking events occur at a certain rate in any stationary random process. In mathematical terms,
a record is an entry in a time series that is larger (upper record) or smaller (lower record) than all previous entries \cite{Arnold1998,Glick1978,Schmittmann1999}. 
If the entries are independent and identically distributed random variables drawn from a continuous probability distribution, the probability $P_n$ to observe a new record after $n$ steps, hereafter referred to as the \textit{record rate}, is simply $P_n = 1/n$, because all $n$ values are equally likely to be the largest. Applying this result to maximal temperatures measured at a specific calendar day over a time span of $n$ years, it follows that the expected number of records per year is $365/n$, i.e. about 12 records for an observation period of 30 years. Remarkably, this prediction is entirely independent of the underlying probability distribution, which may even differ for different calendar days.

\begin{figure}
\onefigure[scale=0.25]{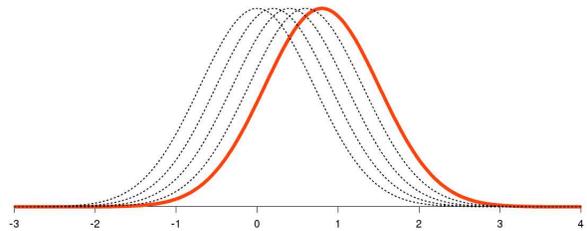}
\caption{(Color online) Schematic of the evolution of the daily temperature distribution under linear drift of the mean.\label{Fig:Gauss}}
\end{figure}

Despite considerable current interest in extreme climate events
\cite{Balling1990,Katz1992,Easterling2000,Schaer2004,Stott,Sabhapandit2007,Nicholls,Meehl2007,Brown2008,Cattiaux,Ballester2010},
the subject of climate records has received relatively little
attention. It is intuitively obvious that an increase in the mean
temperature will lead to an increased occurrence of high temperature
records, but attempts to detect this effect in observational data have
long remained inconclusive
\cite{Bassett1992,Benestad,Benestad2,Redner}. Only very recently an empirical study of
temperature data from the US found a significant effect of warming on the
\textit{relative} occurrence of hot and cold records \cite{Meehl2009}.

\begin{figure}
\onefigure[scale=0.15]{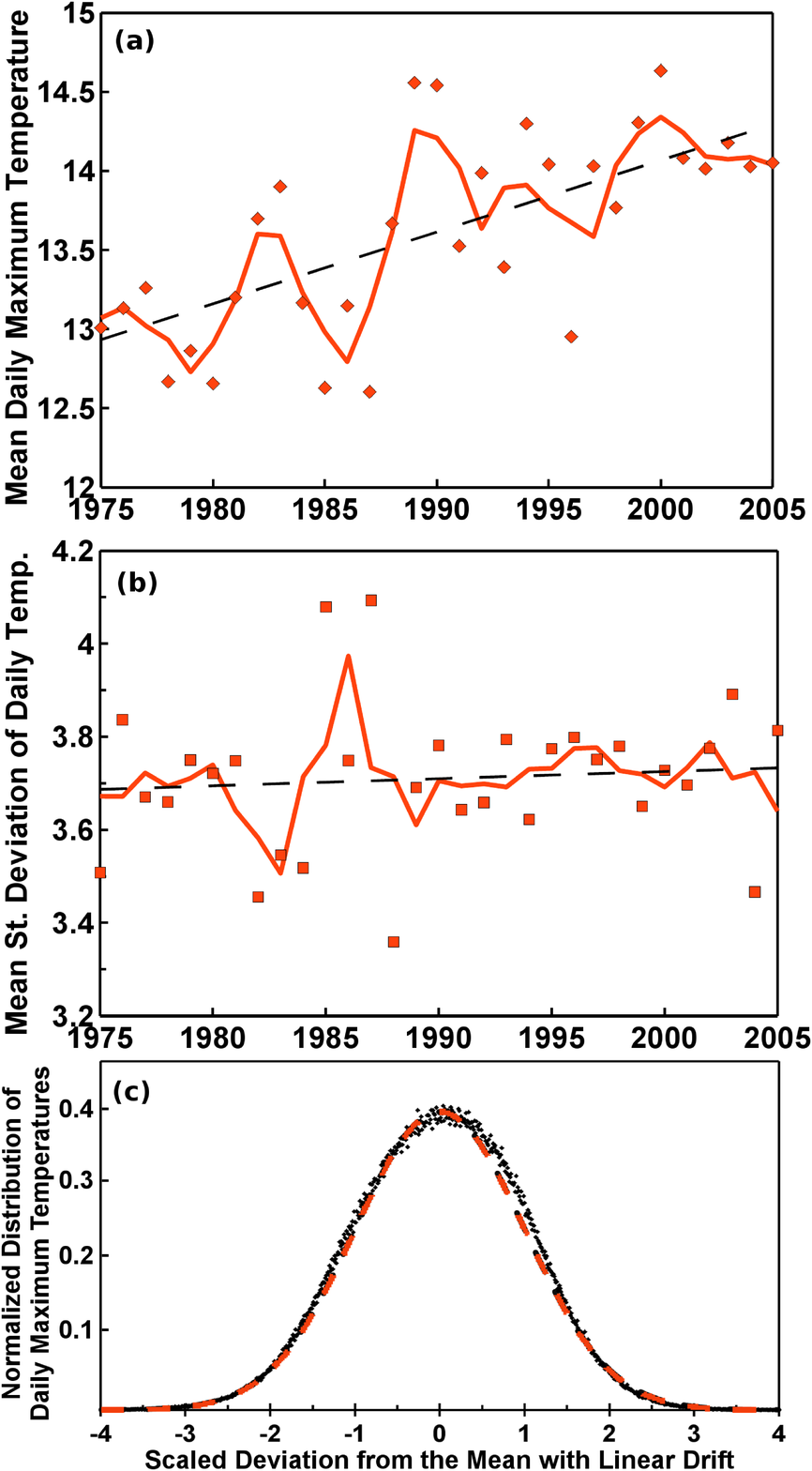} \caption{(Color online) This figure summarizes the behavior of the distribution of daily maximum temperatures for data set EII.
\textbf{(a)} Mean daily maximum temperature. Daily maximum temperatures were averaged over all stations and the entire calendar year. Diamonds show the average of dTmax for individual years and the full line is a sliding 3-year average. The regression line (dashed) shows a clear increase over the last 30 years. \textbf{(b)} Standard  deviation of daily maximal temperatures. To estimate the standard deviation, first a linear regression of dTmax was carried out for each station and each calendar day. The standard deviation for a given year was then computed by averaging the squared deviation of dTmax from the linear fit over all stations and all calendar years. Full line is a 3-year average and the dashed line the result of a linear regression. We find no systematic trend in the standard deviation. \textbf{(c)} Distribution of daily temperatures on individual calendar days. The measurements were detrended and normalized for all time series individually and then accumulated. The dashed line is the probability density of a standard normal distribution.}
\label{Fig:Variance}
\end{figure}

Here we present a detailed analysis of several large data sets of
temperature measurements from both American and European weather
stations, as well as re-analysis data\footnote{The re-analysis approach combines
  meteorological observations from a variety of sources with advanced
  data assimilation techniques in order to create a continuous stream
  of observables on a three-dimensional grid, see \cite{Reanalysis}
  for details.}. We find that the observed increase in the number of high temperature records
(and the corresponding decrease in the low records) is well described by a minimal model
which assumes that the distribution of temperatures measured on a given calendar day is a Gaussian with constant standard deviation $\sigma$
and a mean that increases  linearly in time at rate $v$ (see Fig.\ref{Fig:Gauss}).
This model is consistent with previous findings \cite{USHCN,ECAD,Reanalysis,Redner,IPCC}
and it is supported by our own analysis of the available data sets \cite{Wergen2009}, see Fig.\ref{Fig:Variance} for an example. 
While changes in temperature \textit{variability}
have also been argued to be  important in the generation of extreme
temperature events \cite{Katz1992,Schaer2004}, we have failed to
detect a clear systematic trend in $\sigma$ in the data [
Fig.\ref{Fig:Variance} (\textbf{b})]. Moreover, the increase in the mean supersedes a possible effect on $\sigma$, 
in the sense that the former leads to an asymptotically constant
record rate \cite{Ballerini1985,Ballerini,Borovkov,Franke2010} whereas
the latter only increases the record rate from $1/n$ to $(\ln n)/n$
\cite{Krug2007}. For these reasons we restrict ourselves to the
simplest setting of a temperature distribution of constant shape and
linearly increasing mean. Although temperature fluctuations are well known to display
  long-term correlations \cite{Bunde1998,Bunde2003}, the assumption
  that the daily temperatures are not correlated is justified because
  individual measurements in a time-series are always one year
  apart (see \cite{Redner} and the quantitative discussion below).

\section{Theory}

We begin by deriving an approximate analytic expression for the increase in the record rate $P_n$ caused by a linear drift of the mean. In general,
the record rate for a sequence of independent but not identically distributed random variables $x_n$ is given by \cite{Krug2007}
\begin{eqnarray}\label{Pn_gen}P_n = \int_{-\infty}^{\infty}f_n(x)dx\prod_{k=1}^{n-1}\left(\int_{-\infty}^x dx_k f_k(x_k)\right)\end{eqnarray}
where $f_n(x)$ denotes the probability density at time step $n$.  Here
we consider a drifting distribution of constant shape,  which implies $f_n(x) = f(x-vn)$ with a common density $f(x)$. This reduces (\ref{Pn_gen}) to
\begin{eqnarray}\label{Pn_v}
P_n = \int_{-\infty}^{\infty}f(x)dx\prod_{k=1}^{n-1}\left(\int_{-\infty}^{x+vk} dx_k f(x_k)\right).
\end{eqnarray}
An explicit evaluation of (\ref{Pn_v}) is possible for special choices of $f(x)$, but in general it is only known that $P_n$
converges to a nonzero limit $P^\ast = \lim_{n \to \infty} P_n$ when $v > 0$
\cite{Ballerini1985,Ballerini,Borovkov,Franke2010}. In the climate context the drift speed is expected to be small compared to the standard deviation of the distribution. We therefore expand (\ref{Pn_gen}) to linear order in $v$, which yields
\begin{eqnarray}P_{n} \approx \frac{1}{n} + \frac{v n\left(n-1\right)}{2}\int_{-\infty}^{\infty}dx f^2\left(x\right)F^{n-2}\left(x\right)\end{eqnarray}
where $F\left(x\right)$ is the cumulative distribution function of $f\left(x\right)$. In \cite{Franke2010} the integral in the second term is evaluated for various elementary distributions. For distributions with a power law tail one finds that the correction term decreases for large $n$. 
On the other hand, for distributions that decay faster than exponential, the correction term generally increases with $n$.
In the Gaussian case  of interest here the integral can be evaluated in closed form only for $n = 2$ and 3, with the result 
\begin{equation}
\label{P2P3}
P_2 \approx \frac{1}{2} + \frac{v}{2 \sqrt{\pi} \sigma}, \;\;\; P_3 \approx \frac{1}{3} + \frac{3 v}{4 \sqrt{\pi} \sigma}.
\end{equation}
Using a saddle point approximation and the properties of the Lambert W-functions \cite{Knuth1} to extract the behavior for large $n$ one arrives at the asymptotic expression \cite{Franke2010}
\begin{equation}\label{Pnv}P_n \approx \frac{1}{n} + \frac{v}{\sigma}\frac{2\sqrt{\pi}}{e^2}\sqrt{\ln \left(\frac{n^2}{8\pi}\right)},\end{equation}
which is accurate for $n \geq 7$. For $n = 4,5,6$ the integral can be evaluated numerically. 
For a typical value of $v/\sigma \approx 0.01$ and a time span of 30 years, (\ref{Pnv}) implies an increase of the record rate from $1/30 \approx 0.033$ to 0.042, or an increase in the expected number of record events per year from 12 to 15. In the following this prediction will be compared to empirical temperature data.

\section{European data}

The most comprehensive analysis was carried
out for temperature data obtained within the ECAD project, which
comprises a total of 752 European stations \cite{ECAD}. The data consist of daily recordings for the minimum, mean and maximum temperature, as well as precipitation and snowfall. These stations recorded over time-spans of varying length between less than 10 and more than 200 years. Defective and missing entries were marked in the data sets. We restricted our study to time-series of daily minimum (dTmin) and maximum temperatures (dTmax) which were to at least 95\% reliable. This resulted in two sets of station data, one (set EI) consisting of 43 stations that recorded over the 100 year period 1906-2005, and a second (set EII) containing 187 stations that recorded over the 30 year period 1976-2005. Each station recorded 365 time series, and the results presented below constitute averages over all calendar days and all stations within the respective data set.

Taken together, we thus had roughly 15,000 time series in data set EI
and 68,000 time series in data set EII at our disposal. However, it is
important to note that the effective number of independent time series
is much smaller. The number of independent series is limited by
correlations both in space and in time. The correlations in space
result from the fact that the time series from neighboring stations
are strongly correlated if they are less than 1000 km apart
\cite{Mann1993}. As the spatial distribution of the stations over
Europe was relatively homogeneous, we estimate  an effective number of
12-15 independent stations for the European data. Furthermore,
although daily temperature measurements in subsequent years can be
assumed to be uncorrelated, time series recorded at individual
calendar days that are close to each other are correlated as well. 
Based on the analysis of \cite{Bunde2003} we estimate that
  these correlations extend over a duration of approximately 10 days,
  which implies that the number of independent calendar days is around 36.
We therefore conclude that our analysis  of the European data effectively comprises about 400-500 independent time series.

Figure \ref{Fig:Variance} summarizes the analysis of the distribution
of dTmax for data set EII. The mean maximal temperature is found to
increase at rate $v = 0.047\pm0.003^\circ$C/yr,
while the standard deviation is essentially constant with a mean value
of $\sigma = 3.4 \pm 0.3^\circ$C. 
The detrended temperature fluctuations are Gaussian to a good
approximation. A corresponding analysis for data set EI yields a warming rate of 
$v = 0.0085\pm0.003^\circ$C/yr and the same standard deviation as for
data set EII.

To directly test for correlations between daily
  temperatures, we computed the average two-point
  correlation for subsequent years after
  subtracting the drift and normalizing. For both data sets the
  correlations were found to fluctuate around a small average value
of order $\pm 0.01$ with a standard deviation of order 0.1. These
values are consistent with a power law decay of the form found in \cite{Bunde1998,Bunde2003}.

\begin{figure}
\onefigure[scale=0.6]{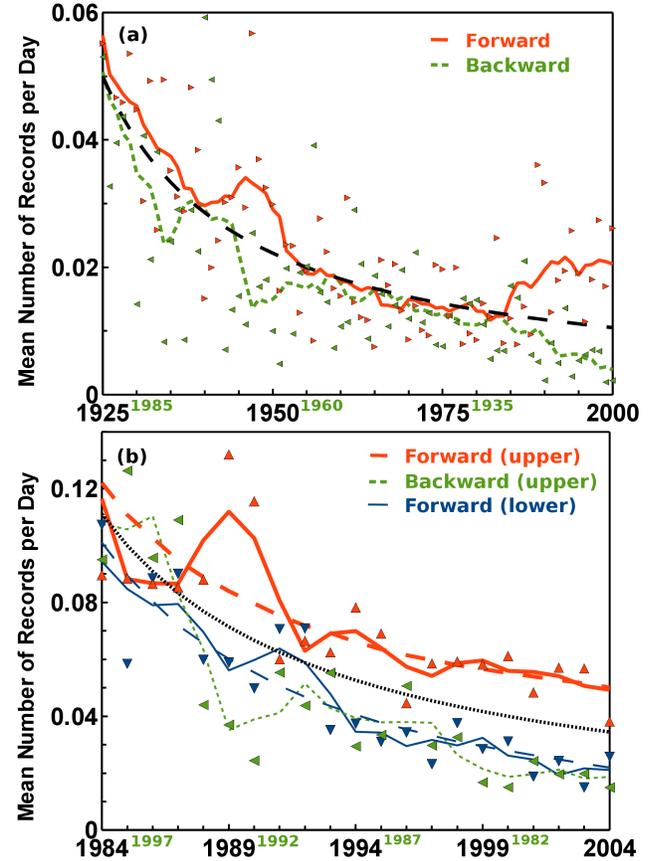}
\caption{(Color online) \textbf{\textbf{(a)}} Record frequencies for data set EI
  (1906-2005). Symbols show the average number of records per calendar
  day in forward ($\triangleright$) and backward ($\triangleleft$) time direction,
  averaged over 365 calendar days and 43 stations. Full and
  short-dashed bold lines were obtained by an additional sliding
  9-year average. Long-dashed bold line shows the prediction $P_n =
  1/n$ for a stationary climate. \textbf{(b)} Record frequencies for
  data set EII (1976-2005). Symbols show the average number of records
  per calendar day for upper ($\bigtriangleup$) and lower
  ($\bigtriangledown$) records in forward direction, and for upper
  records in backward direction ($\triangleleft$). Full lines were obtained by an additional sliding 3-year average. Dotted line shows the prediction $P_n = 1/n$ for a stationary climate, and long-dashed lines show the model predictions. \label{Fig:SetI}} \end{figure}

In Figure \ref{Fig:SetI}\textbf{(a)} we show results of the analysis of temperature records in data set EI. The figure depicts the measured daily record frequency for upper records of dTmax, obtained both from a forward analysis (where a record is the highest value of dTmax since 1906) and from a backward analysis (where years are counted backwards in time and records are defined with respect to the temperature in 2005). According to the prediction (\ref{Pnv}), the forward and backward record rates should lie symmetrically around the record rate $1/n$ of the stationary climate, which is consistent with the displayed data. Throughout the analyzed time span (with the exception of a short period around 1960 in which the climate was effectively cooling)  the forward record frequency lies above the backward record frequency. This shows that the increase in the mean temperature significantly affects the statistics of records. The effect is particularly pronounced during the last two decades, where warming has been most significant (see the discussion of data set EII below). For the year 2005, the measured forward record frequency is about twice as large as expected for a stationary climate. Using the mean warming rate estimated over the entire 100 yr time period, only an enhancement of 40\% is predicted by Eq.(\ref{Pnv}). This shows that the assumption of a constant rate of warming is not a good approximation for data set EI.     

\begin{figure}
\onefigure[scale=0.14]{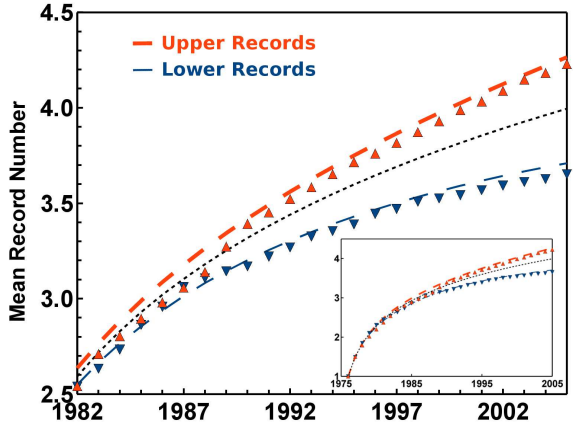}
\caption{(Color
    online) Mean
    record number at European stations (1976-2005). Symbols show the
    average number of upper ($\bigtriangleup$) and lower
    ($\bigtriangledown$) records observed since 1976 at a given
    calendar year in the forward time analysis. Dotted line shows the prediction for a stationary climate, and dashed lines show the prediction for a constant rate of warming. Inset shows the results for the entire time-span from 1976 to 2005.\label{Fig:Mean}}\end{figure}

Figure \ref{Fig:SetI}\textbf{(b)} displays the corresponding results
for data set EII. 
% Owing to the larger number of stations, fluctuations
% are considerably reduced. 
Since the rate of temperature
increase was relatively constant during this time period, 
we find good quantitative agreement between the data and
the model predictions. The agreement is even more striking for the
mean record number displayed in Figure \ref{Fig:Mean}. In a stationary
climate the expected number of records observed over $n$ years is 
\begin{equation}
 \label{Rn}
R_n
= \sum_{k=1}^n \frac{1}{k} \approx \ln n + \gamma
\end{equation}
where $\gamma
\approx 0.5772156...$ is the Euler-Mascheroni constant. For a 30 year
period this amounts to an expected record number of 3.98, which is to
be compared to the observed number 4.24 for the upper records, and
3.66 for the lower records of dTmax. Together Figures \ref{Fig:SetI}
and \ref{Fig:Mean} provide a strong validation of our model. Using our
estimate $v/\sigma = 0.014$ for data set EII, Eq.(\ref{Pnv}) predicts
that the increase 
in mean temperature has increased the rate of record occurrence by
about 40\% over the time period from 1976-2005, 
which implies an additional 5 out of 17 records per year in 2005.

Similar analyses were carried out for upper and lower records of dTmin. We find that the mean record number of dTmin behaves similar to the number of records of dTmax, with 4.32 upper records and only 3.66 lower records. In the backward time analysis we found 3.71 upper records for dTmax and only 3.62 for dTmin. The number of lower records was increased in the backward time analysis, which is in agreement with the results for the upper records. In summary, the number of lower records has decreased in the same manner as the number of upper records has increased (see Fig.\ref{Fig:SetI}\textbf{(b)}).

\begin{figure}
\onefigure[scale=0.06]{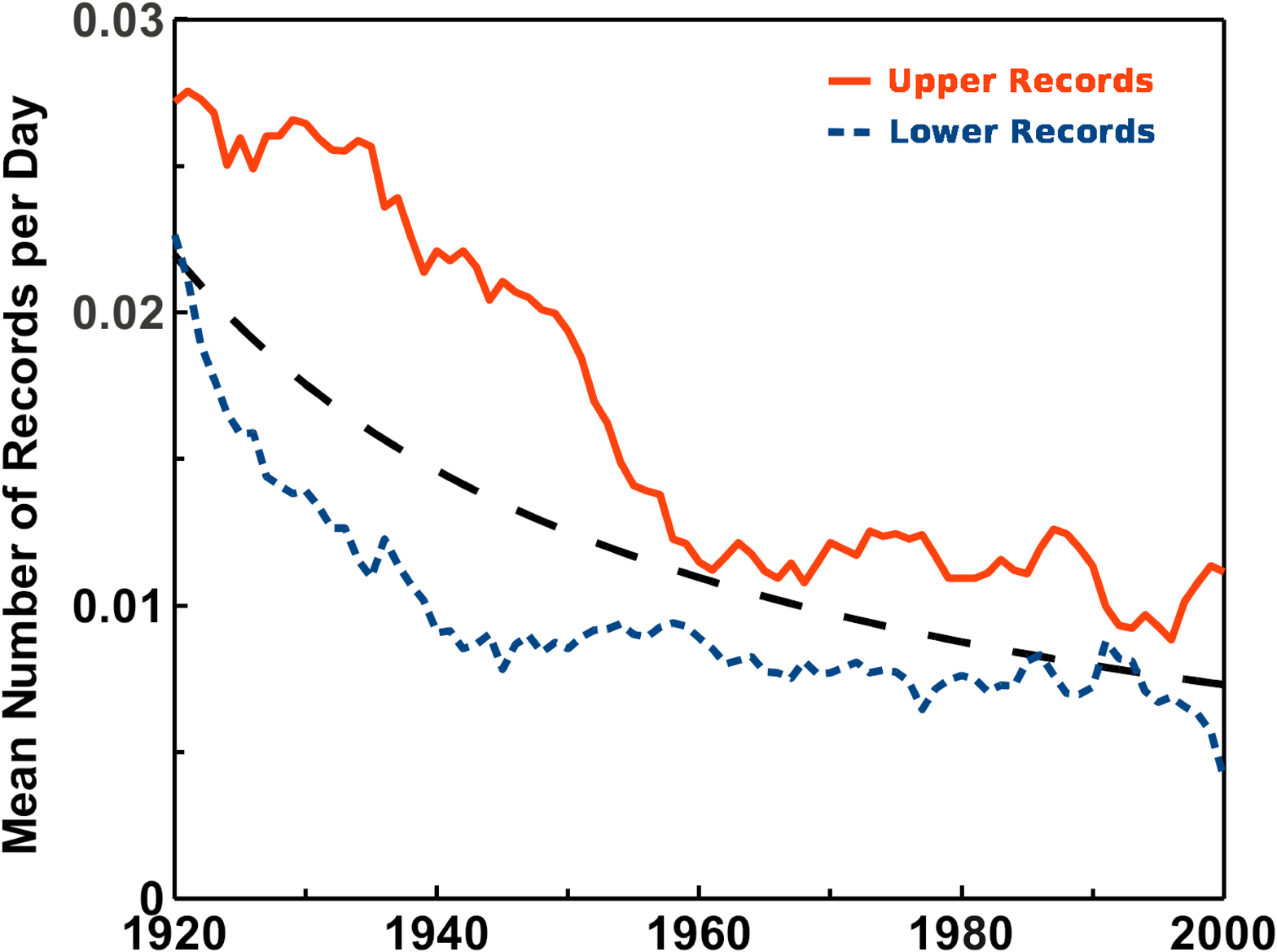}
\caption{(Color online) Record
  frequencies for data set AI (1881-2005). Full line shows the
  average number of upper records per calendar day after a 9-year
  sliding average, short-dashed line shows the corresponding frequency
  for lower records, and long-dashed line shows the prediction $P_n =
  (1-d/\sigma)/n$ for a stationary climate with discrete
  measurements.\label{Fig:American}}
\end{figure}

\section{American data and discreteness effects}
The American data sets were extracted from a
total of 1062 stations \cite{USHCN}. Requiring again a reliability of
at least 95\%, we were left with 10 stations that recorded over the
125 year time span 1881-2005 (data set AI) and 207 stations that
recorded over the 30 year time span 1976-2005 (data set AII). While the 10 stations of data set AI can be assumed to be independent, 
the number of effectively independent time series in data set AII is again much smaller. 

The
result of the record analysis was similar to that performed on the
European data sets, with two important differences. First, owing to
the continental character of the American climate, the standard
deviation $\sigma$ is considerably larger than in Europe, which,
according to Eq.(\ref{Pnv}), implies a weaker effect on the record
rate. For example, for data set AII we estimate a warming rate of $v =
0.025 \pm 0.002^\circ$C/yr and a standard deviation of $\sigma = 4.9
\pm 0.1^\circ$C, which yields a ratio $v/\sigma$ that is only one third of
the value for data set EII. 

Second, the American data were rounded to
full degrees Fahrenheit, whereas the European data were measured in
tenths of degrees Celsius. As a consequence, the probability of ties
is significant in the American data but negligible for the European
data sets. Here we count only \textit{strong} records, which are
broken only by a value that exceeds the current record. To account for
these discreteness effects one computes the probability that a current
record is tied in the $n$th event. For a small unit of discretization
$d\ll\sigma$, this probability is given by $P_n^\mathrm{tie}\approx
d/(\sigma n)$. This leads to the probability for a record event with
discretization as $P_n^d \approx (1 - d/\sigma)/n,$ and summing over
$n$ the reduction of the number of strong records due to ties in a
stationary climate is well described by \cite{Wergen2009} 
\begin{equation}
R_n^d \approx (\ln n +
\gamma)(1 - d/\sigma) + 2d/\sigma. 
\end{equation}
For the American data sets $d =
5/9^\circ\textrm{C} = 0.5555..^\circ$C, which reduces the number of
strong records per day expected in a stationary climate over a 30 year
period from 3.98 to 3.75. In comparison, the observed number of
records in data set AII is equal to 3.86 in the forward analysis and
3.66 in the backward analysis. Again, warming has significantly
increased the number of records, but the effect is less pronounced
than for the European stations. The evolution of record frequencies in
data set AI is shown in Fig.\ref{Fig:American}. Note that a failure to account
for the discreteness effect would lead to an apparent asymmetry between high and low 
records relative to the stationary case. 
Such an asymmetry was observed in the analysis of American temperature data in  
\cite{Meehl2009}, where it was suggested that warming primarily reduces the number
of low temperature records, while the effect on high records is less pronounced.

\begin{figure}
\onefigure[scale=0.35]{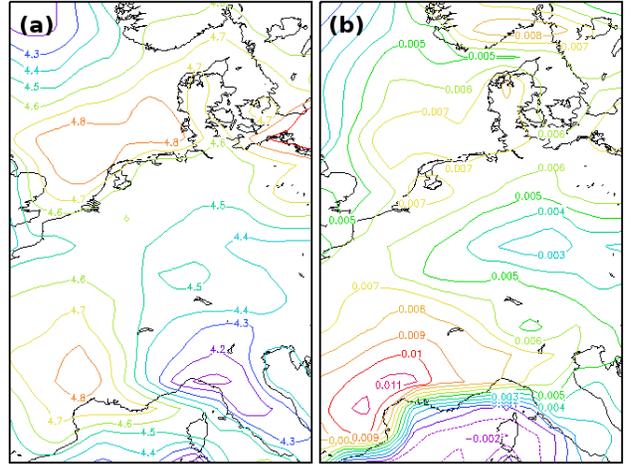}
\caption{(Color online) Spatial distribution of record number and normalized warming rate in central Europe based on re-analysis data (1957-2000). \textbf{(a)} Contour map of the number of records, computed from the 365 time series of daily high temperatures for each point on a rectangular grid of $14 \times 18 = 252$ stations. The expected number of records in a stationary climate is 4.36. \textbf{(b)}  Contour map of the spatial distribution of the rate of warming, normalized by the standard deviation. \label{Fig:recordmap}} \end{figure}

\section{Re-analysis data}
Taken together, the results presented so far show that the increased occurrence of temperature records can be linked quantitatively to the ratio $v/\sigma$ of warming rate and temperature variability. Using the ERA-40 Re-Analysis data \cite{Reanalysis}, we were able to extend this analysis to the spatial distribution of the record rate. The data consist of daily temperature series over the period 1957-2000 for 252 geographic locations in central Europe arranged on a regular grid, covering an area of about $3 \times 10^6 \; \mathrm{km}^2$. For each location the number of upper records of the daily maximal temperature was determined, and the results are shown in the form of a ``record map'' in Fig. \ref{Fig:recordmap}\textbf{(a)}. The comparison with a corresponding map of local values of the ratio $v/\sigma$ in Fig.\ref{Fig:recordmap}\textbf{(b)} shows similar patterns, supporting our conclusion that $v/\sigma$ is a good (if not perfect) predictor for the increased occurrence of records. Interestingly, the two most pronounced islands of high record occurrence ($R_n > 4.8$) in Fig. \ref{Fig:recordmap}\textbf{(a)} are attributed to different mechanisms. One, in southern France, reflects the exceptionally high rate of warming $v$ in this region, whereas the other, over the North Sea, is a consequence of a low temperature variability $\sigma$.

\begin{figure}
\onefigure[scale=0.45]{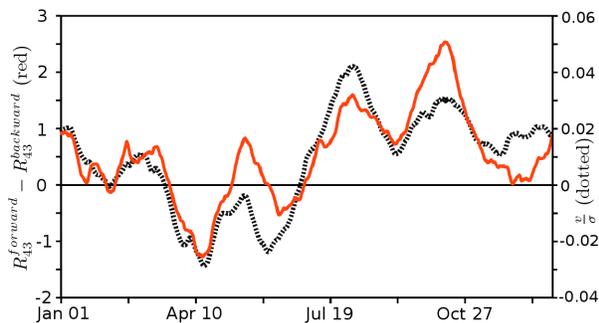}
\caption{(Color online) Seasonal
  distribution of the excess number of temperature records compared to
  the seasonal distribution of $v/\sigma$ in central Europe based on
  re-analysis data (1957-2000). The warming rate $v$ and the standard deviation $\sigma$ for a
  given calendar day was computed as described above in the caption of
  Fig.\ref{Fig:Variance}. The warming rate for a given calendar day is
  the average over all stations of the slope of the corresponding
  linear regression, and the standard deviation is the averaged squared deviation from the linear
  trend over all stations and years. Full line represents the
  difference between the mean number of high temperature records in the
  forward time analysis and the backward time analysis for the entire
  43 year period (left axis). Dotted line gives the seasonal distribution of $v/\sigma$
  (right axis). Both lines were obtained by performing a sliding
  30-day average.\label{Fig:seasonal}}
\end{figure}

An analysis of the seasonal variability of the record events in the (spatially averaged) re-analysis data leads to a similar result. We compared the seasonal distribution of the difference between the mean record numbers in forward and backward time analysis to that of the ratio $v/\sigma$, and found a close match between the two (Fig.\ref{Fig:seasonal}). While the standard deviation shows a clear seasonal pattern with a pronounced maximum in winter, the seasonal variability of the warming rate is rather complicated. As a consequence, a simple seasonal pattern in the rate of record occurrence could not be identified.

\section{Conclusions}
In summary, by combining a simple mathematical
model with extensive data analysis, we have conclusively established
that the current rise in mean temperature significantly affects the
rate of occurrence of new temperature records. While the majority of
the high temperature records observed in Europe at the end of the 30
year period from 1976-2005 would have occurred even in a stationary
climate, the effect of warming is substantial, leading to an
additional 5 out of 17 records per year.  

The key parameter governing the effect of warming on the occurrence of records 
is the ratio $v/\sigma$, and to leading order the change in record rate is linear
in this parameter. 
It is instructive to explore the future 
frequency of record-breaking events under the assumption that
$v/\sigma$ will remain constant. 
The expression (\ref{Pnv}) then predicts that the enhancement of the record
frequency (compared to the expectation $P_n = 1/n$ in a stationary
climate) will continue to increase, up to the point where the expansion underlying
(\ref{Pnv}) breaks down when the two terms become of comparable
magnitude at a time roughly of order $n^\ast \sim \sigma/v$. Beyond this
time the record rate saturates at a constant value $P^\ast$. 
Using our estimate $v/\sigma \approx 0.014$ for data
set EII, we find that $P^\ast \approx 0.033$ for the Gaussian
distribution. This implies that,  towards the end of this century,
daily high temperatures exceeding all values measured since 1976 will
continue to occur in Europe on about 12 days of the year; at the same time the occurrence of low temperature records will essentially cease. 

\acknowledgments
We acknowledge fruitful interactions with Andreas Hense and Sid Redner. This work was supported by DFG within the Bonn Cologne Graduate School of Physics and Astronomy. J.K. is grateful to the Hebrew University of Jerusalem and the Lady Davis Fellowship Trust for their kind hospitality during the completion of this work.

\end{document}